\documentclass[%
 aip,
pof,
 amsmath,amssymb,
preprint,%
]{revtex4-1}

\usepackage{graphicx}
\usepackage{dcolumn}
\usepackage{color,bm}

\usepackage[utf8]{inputenc}
\usepackage[T1]{fontenc}
\usepackage{mathptmx}

\newcommand{\average}[1]{\ensuremath{\langle#1\rangle} } 
\def\R{\ensuremath{\mathrm{Re}}}

\newcommand{\argmax}{\mathop{\rm arg~max}\limits}
\newcommand{\argmin}{\mathop{\rm arg~min}\limits}
\newcommand{\ed}[1]{{#1}}
\newcommand{\ma}[1]{{#1}}
\newcommand{\mb}[1]{{#1}}
\newcommand{\mc}[1]{{#1}}
\newcommand{\ra}[1]{{#1}}
\newcommand{\rb}[1]{{#1}}
\newcommand{\ee}[1]{{#1}}

\begin{document}

\preprint{AIP/123-QED}

\title[Efficient \ed{reinforcement} learning for fluid flow control]
{Efficient \ed{reinforcement} learning \rb{with partially observable} for fluid flow control}

\author{A. Kubo}
\author{M. Shimizu}%
\email{shimizu@me.es.osaka-u.ac.jp}
\affiliation{Graduated School of Engineering Science, 
Osaka University, Toyonaka, 560-0043 Japan}%

\date{\today}

\begin{abstract}
Despite the low \ed{dimensionalities} of dissipative \ed{viscous} fluids, reinforcement learning (RL) requires many observables in fluid control problems. This is because \ed{the} observables are assumed to follow a policy-independent Markov decision process in the RL \ed{framework}. By including policy parameters as arguments of a value function, we construct a consistent algorithm \ed{with} \rb{partially 
observable condition}. Using typical examples of \rb{active flow control}, 
we show that our algorithm is more stable and efficient than the existing RL algorithm\ed{s,} even under a small number of \mc{observables}.
\end{abstract}

\maketitle

\section{INTRODUCTION}

Data-driven \ed{modeling} based on machine 
learning \ee{is currently} more practical than \ee{conventional modeling} 
\ed{based on theories} in \ed{various applications including fluid engineering}. 
Machine learning is used to model, control, and optimize flows from the vast amount 
of data obtained from experiments and simulations. 
For a detailed review of machine\ed{-}learning applications \ed{in} 
fluid mechanics, \ed{refer to} 
Brunton et al. (2020)\cite{brunton2020machine}. 

Reinforcement learning \ra{(RL)}\cite{sutton2018reinforcement} 
is one of machine learning techniques. 
Its algorithm finds the optimal control method that maximizes the total rewards in a given system. 
The learning algorithm\ed{s adopt} \ee{a} trial\ee{-}and\ee{-}error \ed{approach} 
to \ed{determine} the \ee{optimal} control 
from the time-series of states, actions, and rewards. 
Recently, deep reinforcement learning \ra{(DRL)}\cite{franccois2018introduction}, 
the combination \ee{of} deep learning \ee{and RL}, 
has achieved unprecedented performance 
\ed{in artificial intelligence}. 
The first successful examples \ed{of applications include} 
Go, Shogi, and \ed{other} video games\ed{.} 
\ed{These systems significantly outperform 
the best human players} \cite{silver2017mastering}.
\rb{DRL has also been successfully applied 
to industrial applications such as autonomous driving\cite{janai2020computer} 
and controlling of robots\cite{gu2017deep}}. 

\ed{Previous studies have \ee{also} investigated the applicability of} 
\ra{RL} \ed{in} active control, \ra{optimal design, and 
numerical modeling within fluid mechanics}.
Value-based \ee{RL} with Q-learning or SARSA has been applied \ed{in} school movement \ed{optimization} \cite{gazzola2014reinforcement,gazzola2016learning,novati2017synchronisation}, 
\ed{reduction of} drag around cylinders\cite{gueniat2016statistical}, 
and optimal glider control\cite{reddy2016learning,reddy2018glider}. 
Verma et al. (2018) combined deep \ed{long} short-term memory 
with \ra{RL}
to investigate the mechanism of fish school formation \cite{verma2018efficient}. 
In recent years, \ra{DRL} based on \ee{a} \ra{kind of} 
actor-critic algorithm 
has \ed{been} applied to flow control problems. 
\ra{\ee{Many} research\ee{ers} have applied} \ra{DRL} to optimal flow 
control\ee{, including} \ra{fish adaption behaviors\cite{zhu2021numerical}, 
drag reduction around a cylinder (blowing/suction 
\cite{koizumi2018feedback,rabault2019accelerating,tang2020robust,paris2020robust}, rotary oscillations\cite{tokarev2020deep}\ee{,} and 
assistant rotating cylinders\cite{fan2020reinforcement}), } 
\ra{optimal control of Rayleigh--B{\'e}nard 
convection\cite{beintema2020controlling} 
, conjugate heat transfer\cite{hachem2021deep}\ee{,} 
and shape optimization problems\cite{viquerat2021direct}.} 
\ra{Moreover, DRL has been applied to \ee{the} construction of efficient numerical models\ee{,} optimal simulation hyperparameters\cite{pawar2021distributed}\ee{,} 
and optimal large eddy simulation\cite{novati2021automating}.}
\ra{See recent reviews\cite{garnier2019review,rabault2020deep} for more details of \ra{DRL} 
within fluid flow control.}

\ra{Generally, RL algorithms assume 
the time evolution \ed{as} a Markov decision process 
that does not depend on the \ee{means} of control (policy).  
However,  this assumption usually fails \ed{when} learning \ed{is performed} 
in \ee{a} partially observable \ee{condition}.} 
Although fluid systems have infinite degrees of freedom, the number of observations 
\ed{is limited} in \ra{practical} control. 
\ed{In} a dissipative system\ed{,} such as a \ed{viscous} fluid\ed{,} 
the system's final state \ed{(attractor)}  
\ed{requires} \ed{only} a finite number of observables to determine the  state.\cite{temam2012infinite}  
\ee{Because} these finite observables can describe the time evolution, 
their time series \ra{can be regarded as a Markov process}. 
\rb{However, this process depends on policy: 
the way of time evolution among these observables depends on the way of control.  
\ee{Therefore,} in principle we cannot apply existing RL directly to a problem of fluid control with 
partially observable \ee{conditions} 
even if the number of observables is enough to capture the dynamics.} 
In this paper, we describe a framework  
\ra{\ee{that} resolves \ee{the} above frustrating contradiction 
and optimally controls fluid \ed{with a small number of observables},}  
and adapt it to \rb{a few typical active-control} problems  
\rb{as benchmark examples.}.

\section{Algorithm}
\subsection{Deterministic Actor-Critic Algorithm}

\ra{Before introducing our new algorithm,  
we describe the outline of the \ed{existing} 
\ee{RL} algorithm 
only for the on-policy deterministic actor-critic 
method \cite{silver2014deterministic} \ed{to simplify the analysis}.} 
\ra{Most of \ed{previous} studies using 
DRL} for fluid control 
are based on the actor-critic method. 
\ed{Here,} we assume that \ed{our} system evolves \ed{deterministically,} 
and control is determined as a function of \ed{the} state.  
These assumptions may be sufficient for fluid control.  
If exploration \ed{is to be considered}, the algorithm can be changed 
to \ed{an} off-policy type, 
according to \ed{the approach described by} Silver et al. (2014) \cite{silver2014deterministic}. 

\ed{We} consider \ed{a} discrete dynamical system given by the 
\ra{deterministic} map $\bm F$ \ra{(deterministic Markov decision process)}:
\begin{equation}
{\bm s}_{t+1} = {\bm F}({\bm s}_{t},{\bm a_t} ), 
\label{map}
\end{equation}
\ee{where}, $\bm s_t$ is the state of the system at time $t$, 
and $\bm a_t$ is the control to the system (action). 
In the deterministic policy framework\ee{,} the action is a function 
of a state, $\bm a_t=\bm \mu({\bm s}_t)$, 
\ra{where $\bm \mu$ determines the \ee{means} of control 
and is called \ee{the} policy.}
\ed{Generally}, a policy is approximated by a parametric function 
such as $ \bm \mu (\bm s) \simeq \tilde{\bm \mu} (\bm s; \bm \theta)$. 
\ra{Here, \ed{the} parameters $\bm \theta$ are called 
design variables and \ee{are} optimized to maximize the objective function defined below.} 
Hereafter, $ \tilde{\cdot}$ represents a parametric function.
Let the reward be $r_{t} = R({\bm s}_{t}, {\bm a}_{t})$.  
\ed{Then, we} define the value function $V$ and 
the objective function $J$ as follows\ed{:} 
\begin{gather}
V(\bm s; \bm \theta) = \sum_{t=0}^\infty \gamma^{t} 
 \left. r_{t}(\bm s_t, \bm a_t) \right|_{{\bm s}_0=\bm s},  \\
\ra{
J(\bm \theta)=\lim_{T \rightarrow \infty} \frac{1}{T} \sum_{t=0}^{T-1} V(\bm s_t; \bm \theta)
=\int_{\bm s} \rho(\bm s; \bm \theta) V(\bm s; \bm \theta) d \bm s}, 
\label{obj}
\end{gather}
\ee{where}, $\gamma$ is \ed{a} discount rate and \ed{satisfyi\ee{es}} 
$0 \le \gamma<1$\ed{, and} 
$\rho$ represents the probability density in the state space. 
\ra{$V$ means the weighted sum of reward $r_t$, and $J$ is \ee{its} average. 
In the limit of $\gamma \rightarrow 0$, $J$ simply represents the average reward.}
\ee{Optimal} control \ed{determines} the \ee{optimal} design parameters 
$\bm \theta^*$ \ed{that} maximize the objective function \ed{as}  
\begin{gather}
\bm \theta^* = \argmax_{\bm \theta} J(\bm \theta), \\
\tilde{\bm \mu}^*(\bm s) = \tilde{\bm \mu}(\bm s;\bm \theta^*). 
\end{gather}
When the time series of the state is a stationary process 
such as an attractor in a dissipative system, $\bm \theta^*$ \ed{is independent of}  $\gamma$.  
\ed{This is} because $\gamma$ changes $J$ only by a constant multiple.
In the actor-critic method, the objective function $J$ is optimized via the action-value function $Q$ \ed{as}  
\begin{gather}
Q(\bm s, \bm a; \bm \theta) = \sum_{t=0}^\infty \gamma^{t} 
 \left. r_{t}(\bm s_t, \bm a_t) \right|_{{\bm s}_0=\bm s, {\bm a}_0=\bm a}.
\label{QQ}
\end{gather}
\ed{As for} the policy, the action-value function is approximated by the parametric function of $\bm \omega$ such that 
$Q(\bm s, \bm a;\bm \theta)\simeq \tilde{Q}(\bm s, \bm a;\bm \theta,\bm \omega)$. 
At the end of every episode\ed{,} 
the \ed{c}ritic updates $\tilde{Q}$\ed{,} following the SARSA method\ed{, as}  
\begin{gather}
\delta_t = r_{t+1} + \gamma \tilde{Q}(\bm s_{t+1},\bm a_{t+1};\bm \theta_n,\bm \omega_{\ma{n}}) 
- \tilde{Q}(\bm s_{t},\bm a_{t};\bm \theta_n,\bm \omega_{\ma{n}}), \\
\bm \omega_{n+1} = \bm \omega_{n} + 
\alpha_{\omega} \sum_{t \in T_n} \delta_t \nabla_{\omega} 
\tilde{Q}(\bm s_{t},\bm a_{t};\bm \theta_n,\bm \omega_{\ma{n}}), 
\label{critic}
\end{gather}
\ee{where,} the subscript $n$ represents the episode number\ed{,}  
$T_n=\{t_{n1},t_{n2},\cdots ,t_{nN}\}$ \ee{is} the \ed{set of} $N$ samples 
at \mc{$n$-th} episode, \mc{and 
$\alpha_{\omega}$ \ee{is} the learning rate for $\bm \omega$} . 
On the other hand, the actor follows the deterministic policy gradient theorem 
\cite{silver2014deterministic} \ed{given by}  
\begin{equation}
\nabla_{\theta} J(\bm \theta)=
\int_{\bm s} \rho(\bm s;\bm \theta) \nabla_{\theta} \tilde{\bm \mu}(\bm s;\bm \theta)
\nabla_{a} \tilde{Q}(\bm s, \bm a;\bm \theta)|_{\bm a=\tilde{\bm \mu}(\bm s;\bm \theta)} d \bm s, 
\label{pg}
\end{equation}
and updates the design variables by the stochastic gradient \ed{as}  
\begin{gather}
{\bm \theta}_{n+1} = {\bm \theta}_{n} 
+ \alpha_{\theta} \sum_{t \in T_n} 
\nabla_{\theta} \tilde{\bm \mu}(\bm s_t;\bm \theta_n)
\nabla_{a} \tilde{Q}(\bm s_t, \bm a_t;\bm \theta_n), 
\label{actor}
\end{gather}
\ee{where}, \mc{$\alpha_{\theta}$} is the learning rate \mc{for the design variables}. 
\ra{The summation of the second term gives an unbiased estimate of the policy 
gradient \eqref{pg}.}

\ra{From the set of time-series data $(\bm s_t, \bm a_t,r_t)$, the actor-critic method \ed{determines} 
the optimal policy $\bm \tilde{\mu}^{*}$ 
by \ed{repeatedly updating} the action value function 
$\tilde{Q}$ and the policy $\tilde{\bm \mu}$ 
according to the equations \eqref{critic} and \eqref{actor}\ee{,} respectively. 
Note that in the derivation of the stochastic gradient \eqref{actor}\ee{,} 
the dynamics of the system $\bm F$ and the reward $R$ should not depend directly  
on the policy; that is\ee{,} 
$\frac{\partial F_j}{\partial \theta_i} =0$ and $\frac{\partial R}{\partial \theta_i} =0$. 
For details of the derivation, see Silver et al. (2014) \cite{silver2014deterministic} 
and Sutton et al. (1999) \cite{sutton1999policy}.}

\subsection{Algorithm for Fluid Systems} \label{B}

In a dissipative system\ed{,} such as \ra{a viscous fluid}, the solution trajectory of the 
final state \ed{of the system is} often 
in a much lower dimensional space (attractor) 
than \ed{that} of the system \cite{temam2012infinite}.  
Assuming that the \ra{capacity} dimension of the attractor is $D_{\rm e}$, 
the state can be specified in most cases with $2[D_{\rm e}]+1$ observables\cite{temam2012infinite}.
($[\cdot]$ represents \ed{the} Gauss sign.) 
For example, we can determine the state of \ma{a} microorganism 
that swim\ma{s} in water \ed{with} periodic motion 
\ed{using} only three variables.   
Let $\bm o$ be \ma{$N_{o}$} 
observables sufficient to \ra{uniquely determine each moment on the attractor}. 
The transition among the partial observables $\bm o$ is given by the map 
$\bm o_{n+1}=\bm F(\bm o_t, \bm \tilde{\mu}(\bm o_t);\bm \theta)$ \ed{,} 
and \ed{it} depends on \ra{the policy, $\theta$}. 
The \ra{deterministic Markov process} \ed{defined by} the equation \eqref{map} 
\ra{is independent of the policy} 
only when all variables $\bm s$ \ed{are observed}; 
\ra{the state $\bm s$ evolves according to the Navier--Stokes equations with the 
external control\ee{,} which depends only on $\bm s$.} 
\ra{As mentioned in the previous subsection}, 
the policy ($\bm \theta$) independence of 
\ed{the} evolution \ed{of a system} is \ed{a} necessary 
condition of the policy gradient theorem. \cite{sutton1999policy,silver2014deterministic}
\ed{Thus}, even if the attractor is low dimensional, 
many observations are required to update a policy appropriately 
when \ed{using typical} \ee{RL} algorithm\ed{s} for 
fluid control problems. 
In the following \ed{discussion}, we construct the consistent algorithm 
\ed{with} \ra{partially observable}.

The value function can be determined with the minimum number of arguments 
by specifying $\bm o$ and $\bm \theta$. 
Therefore, there exists $\hat{V}$ such that 
\begin{equation}
V(\bm s; \bm \theta) = \hat{V}(\bm o,\bm \theta). 
\end{equation}
We approximate $\hat{V}$ as a parametric function 
for $\bm w$ \ed{as} 
\begin{equation}
\hat{V}(\bm o,\bm \theta) \simeq 
\tilde{V}(\bm o,\bm \theta;\bm w)
= {\bm w}^T {\bm \phi}(\bm o, \bm \theta), 
\label{vw}
\end{equation}
\ee{where}, $\phi_i$ are nonlinear \ra{basis} functions of $\bm o$ and 
$\bm \theta$, which are constructed from 
the time-series data. 
\ra{The construction in this paper will be shown below.}
\ed{We assume that} the policy is a linear function of the observable $\bm o$: 
\begin{gather}
 \tilde{\bm \mu}(\bm o;\bm \theta)=\bm \varTheta \bm o, 
 \label{policy}
\end{gather}
\mc{where $\bm \varTheta$ is matrix representation for $\bm \theta$;   $\varTheta_{ij}=\theta_{(i-1)N_{\rm o}+j}$.}
Arbitrary control becomes possible by increasing the \ed{number} of observables. 
\ed{Otherwise, the use of} nonlinear basis functions enables complex control. 

In our proposed algorithm, the critic and the actor work as follows.
The critic updates $\tilde{V}$ every $N_{\rm e}$ steps of 
a numerical simulation. 
\ma{The unit of this \ee{update} is called \ee{the} \textit{episode}.}
The weight $\bm w_n$ of $\tilde{V}$ 
at the end of \ee{the} \mc{$n$-th} episode is calculated as follows: 
\begin{gather}
\delta_t = \sum_{k=0}^{N_s-1} \gamma^k r_{t+k} + \gamma^{N_s} \tilde{V}(\bm o_{t+N_s},\bm \theta_{t+N_s}) 
-  \tilde{V}(\bm o_{t},\bm \theta_t),  \\
\bm w_n = \argmin_{\bm w} \sum_{t \in T_n} \delta_t^2, 
\label{wn}
\end{gather}
\ee{where,} $\delta_t$ represents the $N_{s}$ step TD error. 
$T_n=\{(n-M)N_{\rm e}, (n-M)N_{\rm e}+1,\cdots,nN_{\rm e}-1\}$
is the set of steps in the last $M$ episodes. 
\ed{The} basis functions ${\bm \phi}$ are constructed from the time series 
$\{(\bm o_t,\bm \theta_t)| t \in T_n \}$. 
We use \ee{the} RBF sampler\cite{rahimi2009weighted} of the package 
scikit-learn\cite{pedregosa2011scikit} for \ed{the} basis functions. 
Then\ed{,} the actor calculates the stochastic gradient $\bm g_n$ at 
the $n$-\ed{th} episode by \ed{the} linear regression algorithm: 
 \begin{gather}
 \bm g_n, b_n = \argmin_{\bm g, b}  \sum_{t \in T_n} (\bm g^T \bm \theta_t + b 
- \tilde{V}(\bm o_{t},\bm \theta_t))^2. 
\label{an}
\end{gather}
\ed{In} both minimization problems, \ed{given by equations} 
\eqref{wn} and \eqref{an}, 
the singular value decomposition is used to \ed{avoid} singularit\ed{ies} 
of linear equations. 
Using \ed{the} gradient, $\bm g_n$, 
the policy is updated \ed{at each} time step in episode $n$ as follows: 
\begin{gather}
{\bm \theta}_{t+1} = {\bm \theta}_{t} 
+ \alpha \bm g_n, 
\label{actor2}
\end{gather}
\ee{where}, $\alpha$ is the learning rate.
In the limit $\alpha \rightarrow 0$, 
the orbit $\bm o_t$ changes in a quasi-static manner, and 
$\tilde{V}$ is defined on the attractor.

\section{Benchmark results}

\subsection{Drag reduction \ed{through} blowing and suction}
We apply 
the proposed algorithm to the active flow control 
problem \cite{rabault2019artificial} 
\ed{to validate its effectiveness}. 
The goal of this problem is to minimize the magnitude of \ed{the} drag acting on a cylinder 
by controlling the mass flow rate of two synthetic jets symmetrically located 
on the top and bottom of \ed{a} cylinder immersed in a two-dimensional incompressible \ed{fluid}.

The flow configuration and optimization problem 
\ed{have been adopted from} Rabault et al. (2019) \cite{rabault2019artificial}. 
\ra{
The difference from previous settings is that (i) the policy function is a linear function \eqref{policy} rather than a nonlinear function constructed with a neural network, and (ii) the policy parameters are updated with the proposed algorithm in \S \ref{B} instead of \ee{p}roximal policy optimization (PPO) \cite{schulman2017proximal}.

The flow in this system follows the governing non-dimensional equations given by
\begin{eqnarray}
&&\nabla \cdot {\bm u} = 0 ~~~~~~~~~~~~~~~~~~~~~~~~~~~~~~~~~~~~~~~~~~~~{\rm in}~ \varOmega,  \\
&&\frac{\partial \bm u}{\partial t} + ({\bm u} \cdot \nabla) {\bm u} = -\nabla p + \frac{1}{\R} \Delta {\bm u}  ~~~~~~~{\rm in}~ \varOmega,   \\
&&{\bm u} = \frac{6}{4.1^2} \left(\frac{4.1}{2}-y \right) \left(\frac{4.1}{2}+y \right) {\bm e}_x ~~~{\rm on}~ \varGamma_{\rm in}, \\
&&-p {\bm n} + \frac{1}{\R} \nabla {\bm u} {\bm n} = {\bm 0} ~~~~~~~~~~~~~~~~~~~~~~~~~~{\rm on}~ \varGamma_{\rm out},  \\
&&{\bm u} = {\bm 0}  ~~~~~~~~~~~~~~~~~~~~~~~~~~~~~~~~~~~~~~~~~~~~~~~~~{\rm on}~ \varGamma_{\rm wall}, \\
&&{\bm u} = Q_{1} \frac{{\rm \pi}}{10} \cos \left( \frac{{\rm \pi}}{10} (\varphi - 90) \right) {\bm e}_r  ~~~~~~~~{\rm on}~ \varGamma_{1},  \\
&&{\bm u} = Q_{2} \frac{{\rm \pi}}{10} \cos \left( \frac{{\rm \pi}}{10} (\varphi - 270) \right) {\bm e}_r  ~~~~~~~{\rm on}~ \varGamma_{2},
\end{eqnarray}
where $\varOmega$ is a whole computational domain; $\varGamma_{\rm in}$ is an inflow boundary; $\varGamma_{\rm out}$ is an outflow boundary; $\varGamma_{\rm wall}$ is a no-slip boundary\ee{---}that is, top and bottom walls and the solid walls of the cylinder except for the jet parts $\varGamma_i, i=1,2$; and ${\bm e}_x$ and ${\bm e}_r$ denote unit vectors in the streamwise ($x$-) and radius direction, respectively.
It should be noted that the two jets (jet~1 and jet~2) are implemented on the cylinder as angles $\varphi=90^\circ$ and $270^\circ$, \ee{respectively,} and the total mass flow rate injected by them is set to null\ee{---}i.e.\ee{,} $Q_1 + Q_2 = 0$ and $|Q_1| \le 0.1$, where $\varphi$ is an angle from the streamwise ($x$-) direction.
The Reynold\ee{s} number $\R$ based on the bulk mean streamwise velocity  and cylinder diameter is set to 100.

Pressures on the probes located in the vicinity of the cylinder and near the wake are observed.
In the previous study, learning was performed in three cases of 5, 11, and 151 pressure probes, but in this study, only two case\ee{s} of 5 and 11 are conducted.

The reward function is defined by
\begin{equation}
r_t = -\left< C_D \right>_T - 0.2 |\left< C_L \right>_T|,
\end{equation}
where $\left< \cdot \right>_T$ denotes the sliding average back in time over a duration corresponding to one vortex shedding cycle.
$C_D$ and $C_L$ are the drag and lift coefficient\ee{s} on the cylinder ($\varGamma_{\rm {cylinder}}$) given by 
\begin{eqnarray}
C_D &=&  \int_{\varGamma_{\rm {cylinder}}} 2({\bm \sigma} {\bm n}) \cdot {\bm e}_x d \varGamma, \label{C_D}\\
C_L &=&  \int_{\varGamma_{\rm {cylinder}}} 2({\bm \sigma} {\bm n}) \cdot {\bm e}_y d \varGamma, \label{C_L}
\end{eqnarray}
respectively\ee{,} 
\ee{where}, ${\bm \sigma}$ is the Cauchy stress tensor and ${\bm e}_y$ is a unit vector in the wall-normal ($y$-) direction.

The trick to help the agent learning is implemented \cite{rabault2019artificial}: the mass flow rate by two jets is obtained as $Q_{i,t+1} = Q_{i,t} + 0.1(0.1\tanh(a_{i,t}) - Q_{i,t})$, where $Q_{i,t}$ and $a_{i,t}$ are the mass flow rate and the action
\ee{, respectively,} at time $t$ of jet~$i$.
}

We use \ma{their} open\ee{-}source code and replace only the \ee{RL} 
part of that code \mc{with our proposed algorithm above}.  
The parameters of \ee{RL} 
\ed{are as follows:} $ N_ {\rm s} = 200, N_ {\rm e} = 4000, M = 20 $
, and  \mc{$\alpha = 0.0002(1-\frac{n}{500+M}) $}. 
\ra{
4000 steps correspond to 20 dimensionless times (\ee{the} numerical time step is set to $5 \times 10^{-3}$) and approximately \ee{six} vortex shedding periods in the case without blowing/suction control.
}
\mc{In the first $M$ episodes ($1 \le n \le M$), the policy gradient $\bm g_n$ 
	is given randomly.} 

\begin{figure}[h]
	\includegraphics[width=8.5cm]{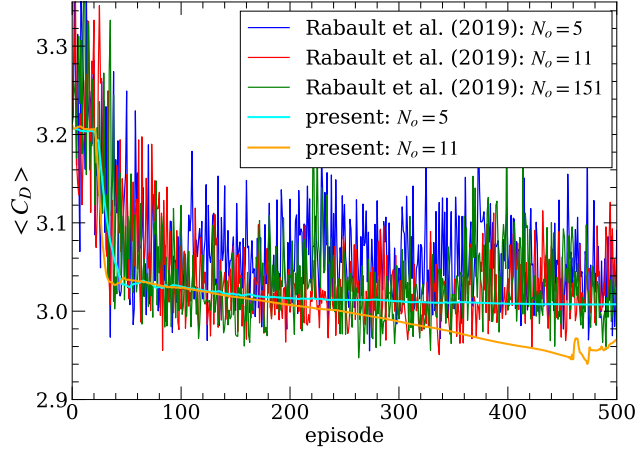}
	\caption{Learning history of the blowing/suction problem.
		$\average{C_D}$ is the time-averaged drag coefficient $C_D$ within 
		one episode.}
	\label{BS1}
\end{figure}

\begin{figure}[h]
	\includegraphics[width=8.5cm]{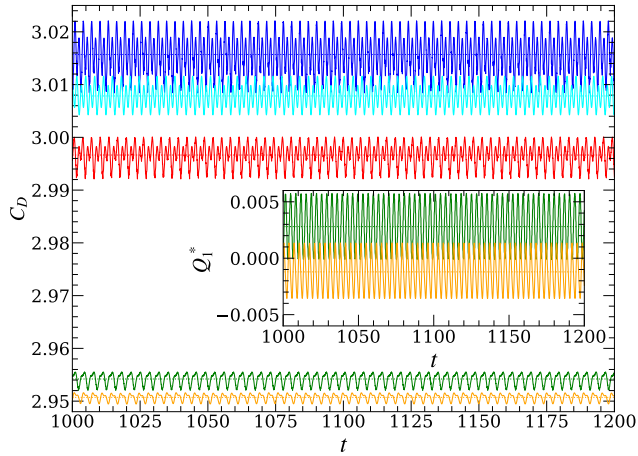}
	\caption{Time variation of the drag coefficient $C_D$ (lines). 
		The dotted lines represent the time averaged value for each case. 
		\mc{Each color represents the same case as in Fig. 1.} 
		\mc{The inset shows \ee{the} time series of 
\rb{normalized mass flow rate} $Q_1^*$ \cite{rabault2019artificial}. }}
	\label{BS2}
\end{figure}

Figure \ref{BS1} shows the learning history, where $N_o$ denotes the number of observables 
(pressure probes in this problem).
From this figure, it can be seen that the drag coefficient $C_D$ decreases more stably \ed{when} learning 
using the proposed algorithm \ed{compared to that}  
in the previous stud\ed{ies} \mc{with stochastic policies.} 
\ma{This is mainly because the optimization process 
	in the proposed algorithm is deterministic, and 
	the trajectory remains close to an attractor 
	for each policy parameter $\bm \theta$.}
The time variation of $C_D$ obtained according to the optimal policy \ed{for} each case is shown in Fig.\ref{BS2}. 
\mc{In the case of $N_{\rm o}=11$, we consider $\bm \theta$ at \ee{the} 446th episode, 
	which is approximately 10 episodes before the local minimum of the objective 
	function, as the optimal policy. Immediately after this local minimum, 
	the flow becomes unstable, and the stochastic gradient fails to increase the 
	objective function.} 
In \ee{the} statistical stationary state, after a sufficiently long term from the initial condition, our linear policy with $N_o=11$ has achieved slightly larger 
drag reduction 
compared to \ed{a} previous study \cite{rabault2019artificial} adopting 
\ed{a} large-scale neural network policy function with $N_o=151$.

\subsection{Optimization of propulsion direction}

\begin{figure}[h]
	\includegraphics[width=7.5cm]{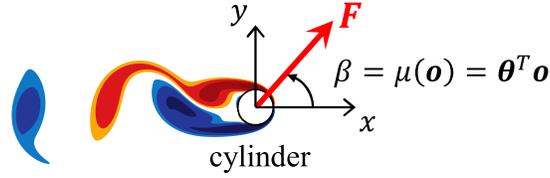}
\caption{Configuration of the cylinder forcing problem. 
\mb{Reinforcement learning optimizes the policy $\beta$, 
the direction of forcing, to m\ee{a}ximize \ee{the} average speed 
of the cylinder $\average{V_x}$.}}
	\label{cylinder}
\end{figure}

For the second demonstration, 
we consider a cylinder driven by a constant\ed{-}magnitude external 
force $ \bm F $ in a 2D incompressible flow field (Fig.~\ref{cylinder}).
The Reynolds number $\R$ based on the external force is set to 100\ed{,} 
and \ed{thus} a vortex street 
is observed in \ed{the} wake of the cylinder, leading to 
\ed{suppression of} the cylinder movement.
The cylinder is a rigid body \ma{whose} mass density \ra{$\rho_c$}
\ma{is two times larger than that of} the fluid \ra{$\rho_f$} \ra{\ee{---i.e.,} 
$\rho_r = \rho_c/\rho_f = 2$}\ee{---} and its rotational motion is ignored for simplicity.
The goal here is to maximize the time-averaged velocity $ \average{V_x} $ of the cylinder 
by controlling the direction $ \beta $ of the external force.

\ra{
The flow $\bm{u}$ and cylinder velocity $\bm{V}$($=(V_x,V_y)^{ T}$) in this system follow the governing non-dimensional equations given by
\begin{eqnarray}
&&\nabla \cdot {\bm u} = 0,  ~~~~~~~~~~~~~~~~~~~~~~~~~~~~~~~~~~~~~~~~~~~~~~~~~~~~~{\rm in}~ \varOmega_{\rm f},   \\
&&\frac{\partial \bm u}{\partial t} + ({\bm u} \cdot \nabla) {\bm u} = -\nabla p + \frac{1}{\R} \Delta {\bm u} + {\bm f}_{\rm{damp}}, ~~{\rm in}~ \varOmega_{\rm f},   \\
&&{\bm u} (x,y) = {\bm u} (x+50,y) = {\bm u} (x,y+10)~~~~~~~~~~{\rm in}~ \varOmega_{\rm f},\\ 
&&\rho_r \frac{{\rm d} {\bm V}}{{\rm d} t} = \int_{\partial \varOmega_{\rm c}} {\bm \sigma} {\bm n} {\rm d} {\bm x} + (\cos \beta, \sin \beta)^{T}, 
\end{eqnarray}
where $\varOmega_{\rm f}$ is fluid domain.
Periodic boundary conditions are imposed on the $ x $- and $ y $-directions, and their periods are taken as $ 50 $ and $ 10 $, respectively, but the flow is stopped by the damping term ${\bm f}_{\rm{damp}}$ in region 40 downstream from the cylinder to eliminate the inflowing vortex.

DNS is performed in a noninertial system with \ee{a} fixed cylinder, and the immersed boundary method is employed \ee{for} the object representation \cite {uhlmann2005immersed}.
The uniform staggered grids are employed with the number of grid points being $500 \times 50 $ in the $x$- and $y$-directions and the numerical time step is $10^{-2}$.
}

\ed{The} observables \ed{are} the $y$-component of velocity $V_y$  of 
the cylinder, lift $L \ra{= C_L/2}$ \ra{(see \eqref{C_L})} and 
time\ee{-}delayed values 
of these two quantities: $\bm o_t=(V_{y,t},L_{t},V_{y,t-N_{\rm s}},L_{t-N_{\rm s}},V_{y,t-2N_{\rm s}},
L_{t-2N_{\rm s}},\cdots) ^{T}$.
The reward is defined as $r_t = V_{x,t} $.
The parameters of \ee{RL} 
\ed{are} $ N_ {\rm s} = \mc{100}, N_ {\rm e} = \mc{5000}, M = 20 $, 
and \mc{$\alpha = 10^{-4-[\frac{n}{500+M}]} )$
($[\cdot]$ represents the Gauss sign)}. 
\ra{5000 steps correspond to approximately \ee{four} vortex shedding periods in the case when external forces 
are applied from a constant angle $\beta = 0$.}

\begin{figure}[h]
	\includegraphics[width=8.5cm]{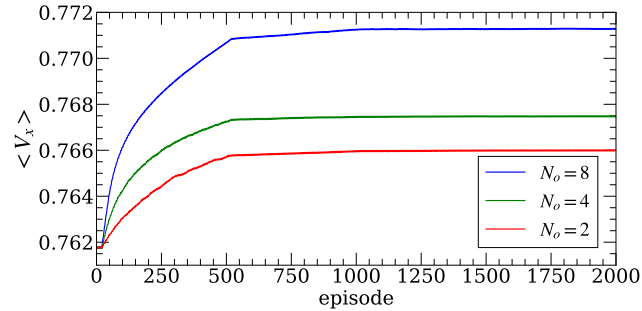}
\caption{Learning history of the cylinder forcing problem.
$\average{V_x}$ is the time-averaged $V_x$ \ed{per} 
episode.}
	\label{force1}
\end{figure}

\begin{figure}[h]
	\includegraphics[width=8.5cm]{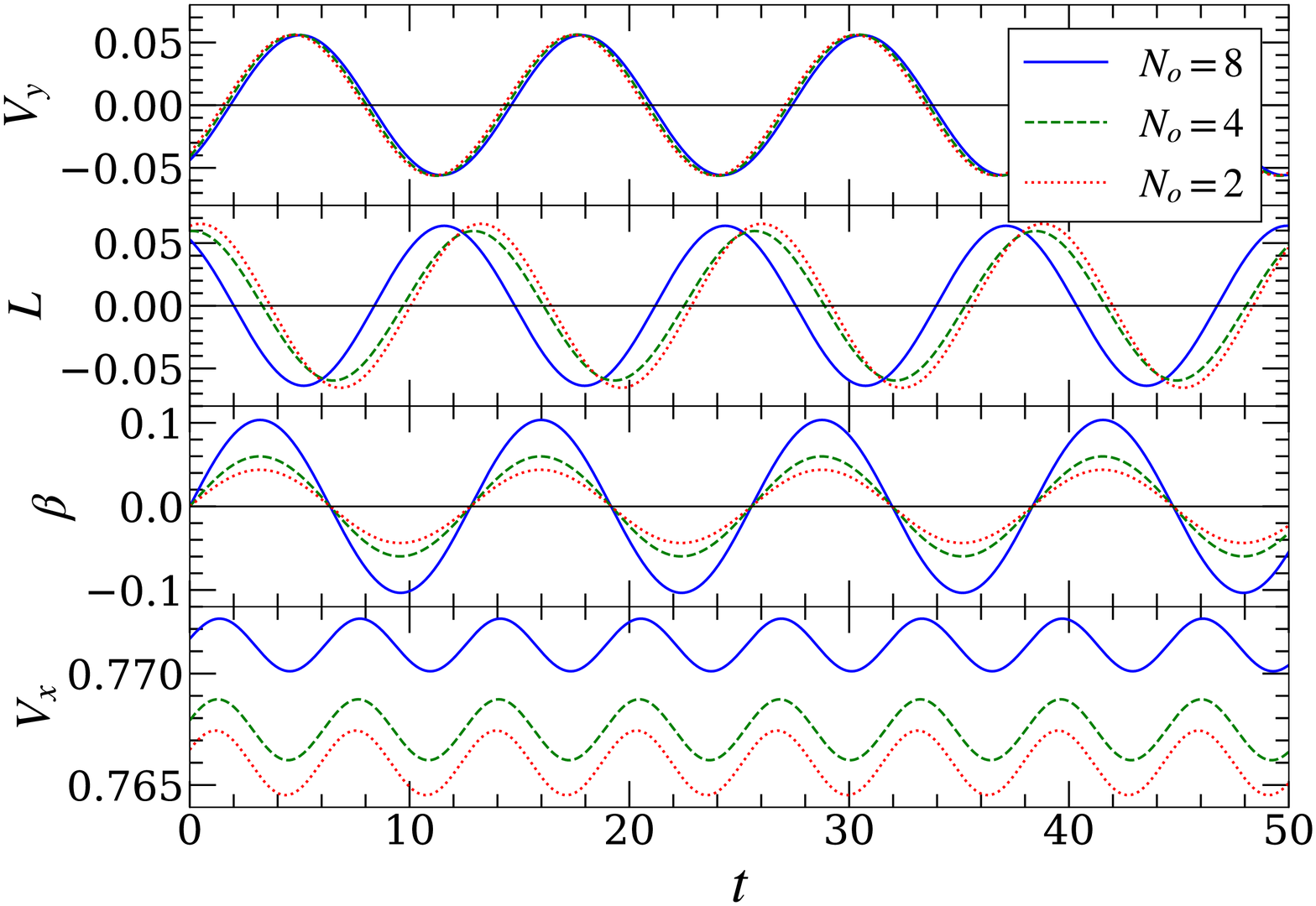}
\caption{\mb{Trajectories of the observables $V_y,L$
, the action $\beta$, and the reword $V_x$ under  
the optimal policies for differ\ee{e}nt numbers of obse\ee{r}vables $N_{\rm o}$.}}
	\label{obs}
\end{figure}

Figure \ref{force1} shows the learning history in the cases 
of $N_o=2$, 4 and 8.
As the number of observables increases, \ed{the} objective function \ed{increases rapidly}. 
The performance with $N_o=8$ is improved by 
\ed{approximately} 1\% compared to the case when external forces 
are applied \ed{from} a constant angle.
\mb{Fig. \ref{obs} represents difference \ee{in} trajectories among these 
three cases.  }  
In this control problem, the flow \ed{becomes} 
unstable when the cylinder velocity is relatively high.  
\ed{Thus,} for successful \ma{and stable} learning, \mc{we \ee{must} adjust} the learning rate $\alpha$ to be small \mc{around an optimal design variable}.


\subsection{\ra{Speed maximization of a swimming hinge}}\label{hinge}

\ra{
For the third demonstration,\rb{we consider a swimming hinge with 
infinitesimal thickness}, 
driven by an external torque $T \le T_{\rm max}$ in a 2D incompressible flow field (Fig.~\ref{hinge}).
\rb{The torque acts only on the plate~1, 
and the plate~2 is allowed to move only in the $ x $-direction}.

The Reynolds number $\R$ based on the maximum torque is set to 100, and thus, a vortex street is observed in the wake of the object, leading to suppression of the movement.
The object is a rigid body \ma{whose} mass density 
is 1.1 times larger than that of the fluid, that is $\rho_r= 1.1$.
The goal here is to maximize the time-averaged velocity $ -\average{x_h} $ of the object 
by controlling the magnitude of the external torque $T=\tanh (\tilde {\mu})+0.1\sin(2 {\rm \pi} t/5)$.
The second term of this torque is introduced 
\rb{to escape stably from null state};   
the speed of the object driven only by this term is \rb{approximately} 0.02, 
which is small enough compared with 
the speed under the following optimum control $\sim 0.5$. 

The \rb{velocity fields} $\bm{u}$ and \rb{the positions of the join point} ${x}_h$ in this system follow the governing non-dimensional equations given by
\begin{gather}
~~~~~~~\nabla \cdot {\bm u} = 0 ~~~~~~~~~~~~~~~~~~~~~~~
~~~~~~~~~~~~{\rm in}~ \varOmega_{\rm f},  \\
\frac{\partial {\bm u}}{\partial t} + ({\bm u} \cdot \nabla ){\bm u} = - \nabla p + \frac{1}{\R} \Delta {\bm u}~~~~~{\rm in}~\varOmega_{\rm f}, \\
{\bm u} = (\dot{x}_h - d \dot{\beta} \sin\beta,0)^T
~~~~~~~~~~~~~~~~~~~{\rm in}~ \varOmega_{\rm 1},  \\
2 \rho_r \ddot{x}_h - \frac{\rho_r}{2} (\ddot{\beta} \sin \beta + \dot{\beta}^2 \cos \beta) = G_x,   \\
\frac{\rho_r}{3} \ddot{\beta} - \frac{\rho_r}{2} \ddot{x}_h \sin \beta = K + T, 
\end{gather}
where $\varOmega_{\rm f}$ and $\varOmega_{i}$ are fluid and 
plate $i$, respectively. 
\rb{Dot notation represents derivative with respect to time.}
$d$ is the one-dimensional coordinate on the flat 
plate~1 with the hinge as the origin, and $\beta$ is 
the angle formed by the flat plate~1 from the $x$-direction.
${\bm G}$ \rb{denotes the sum of the force acting on 
the hinge from the viscous stress,} 
$\bm G=\int_{\partial \varOmega_{1}+\partial \varOmega_{2}} 
{\bm \sigma} {\bm n} ~d {\bm x}$ 
and $K$ the sum of the torque acting on 
the rotational plate~1 from the viscous stress, $K=\int_{\partial \varOmega_{1}} 
[\rb{{\bm x}} \times ({\bm \sigma} {\bm n})]_z ~d {\bm x}$.

DNS is performed in a non-inertial system \rb{in which the joint point is 
fixed at the origin}, and the immersed boundary method is employed to 
\rb{solve fluid--structure interaction}\cite {uhlmann2005immersed}.
The uniform staggered grids are employed with the number of grid points being $400 \times 400 $ in the $x$- and $y$-directions and the numerical time step is $5 \times 10^{-3}$.
	
\ed{The} observables \ed{are} the angle and angular velocity of flat plate~1 and time delayed values 
of these two quantities: $\bm o_t=(\beta_t,\dot{\beta}_{t},\beta_{t-N_{\rm s}},\dot{\beta}_{t-N_{\rm s}},\beta_{t-2N_{\rm s}},
\dot{\beta}_{t-2N_{\rm s}},\cdots) ^{T}$.
The reward is defined as $r_t = -\dot{x}_{h,t} $.
The parameters of reinforcement learning 
\ed{are} $ N_ {\rm s} = \mc{200}, N_ {\rm e} = \mc{4000}, M = 10 $, 
and \mc{$\alpha = {\rm Max}(10^{-5-[\frac{n}{500+M}]},10^{-7})$}. 
	
Figure \ref{hinge_result} shows the learning history in the cases 
of $N_o=2$, 4 and 8.
\rb{As in the example of the previous subsection,}  
the objective function increases rapidly as the number of observables increases.
\rb{However, the dependence on the number of observables on the optimal values is larger 
than the previous problem, and the objective function $-\average{\dot{x}_h}$ 
doesn't improve at all for $N_o=2$.}
	
\begin{figure}[h]
	\includegraphics[width=7.5cm]{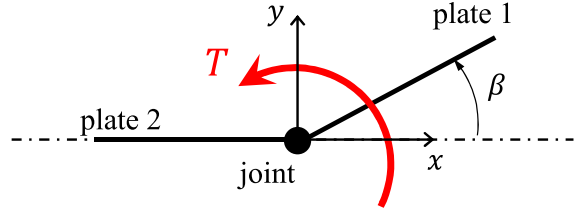}
	\caption{Configuration of the swimming problem. 
\rb{The hinge is driven by the external torque $T$ acting on the plate~1.  
		Reinforcement learning optimizes $T$ to maximize the average speed 
			of the hinge $-\average{\dot{x}_h}$.}}
	\label{hinge}
\end{figure}
\begin{figure}[h]
	\includegraphics[width=8.5cm]{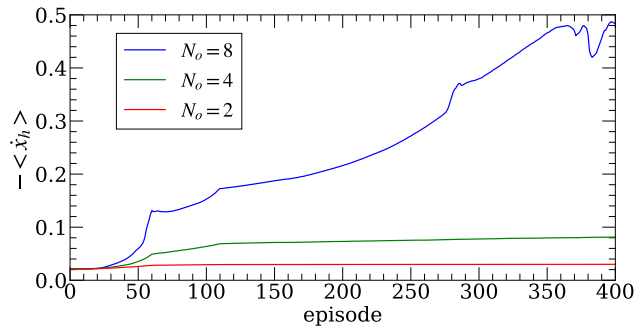}
	\caption{Learning history of the swimming problem 
\rb{for different number of observables}.
		$\average{\dot{x}_h}$ is the time-averaged $\dot{x}_h$ \rb{within an}  
		episode.}
	\label{hinge_result}
\end{figure}
	
}

\section{Conclusion}

Existing reinforcement 
learning algorithms for fluid control
are inefficient under a small number of observables even if the flow is laminar.
By incorporating the low-dimensional nature of the dissipative system into the learning algorithm, 
we resolved this problem and presented a framework for reinforcement learning that stably optimizes 
the policy for a small number of 
observables.
\rb{In the practical aplication of a learning process in a fluid, 
a learning agent often cannot know any information about flow state 
except its own motion. 
For example, a swimming organism can learn how to swim underwater only 
from the time series of its swimming speed and its motion of moving parts as the 
example in the subsection III-C. In such a situation, the algorithm in this paper can efficiently find for the optimum control method. However, the present algorithm cannot handle a flow with large external noise; an inflow with white noise causes the attractor dimension  infinite.} 

As demonstrated in previous studies, the orbit of flow hardly reaches the attractor during the learning process in the probabilistic policy framework. This leads to inefficient optimization \mc{around} the attractor.   
Moreover, the deterministic policy described in this paper enables stable optimization on an attractor, 
and it is not necessary to learn the processes repeatedly. 

Although the proposed algorithm is simple, the value function can be approximated using a neural network 
as in deep reinforcement learning, instead of the generalized linear model of form (11). 
A more complicated regression model may be applied instead of the linear regression model of (15).
In addition, although on-policy-type learning was used in this study, 
the proposed algorithm can be easily modified to the off-policy type to emphasize exploration.

\begin{acknowledgments}
This work was supported by JSPS KAKENHI Grant Number JP17K14588.
Simulations were performed on the ``Plasma Simulator'' 
(NEC SX-Aurora TSUBASA) of NIFS with support 
\ed{from} the NIFS Collaboration Research program (NIFS19KNSS124). 
\end{acknowledgments}

\section*{DATA AVAILABILITY}
The source codes that generate the results of this study are openly available in GitHub repository \cite{KS2021git}.

%

\end{document}